\documentclass[apj]{emulateapj}

\def\ltsima{$\; \buildrel < \over \sim \;$}
\def\simlt{\lower.5ex\hbox{\ltsima}}
\def\gtsima{$\; \buildrel > \over \sim \;$}
\def\simgt{\lower.5ex\hbox{\gtsima}}
\def\kms{{\rm\,km\,s^{-1}}}

\def\kpc{{\rm\,kpc}}

\def\pc{{\rm\,pc}}

\def\deg{^\circ}

\def\s{\ifmmode \widetilde \else \~\fi}
\def\={\overline}

\def\spose#1{\hbox to 0pt{#1\hss}}

\def\lta{\mathrel{\spose{\lower 3pt\hbox{$\mathchar"218$}}
     \raise 2.0pt\hbox{$\mathchar"13C$}}}
\def\gta{\mathrel{\spose{\lower 3pt\hbox{$\mathchar"218$}}
     \raise 2.0pt\hbox{$\mathchar"13E$}}}
\def\Dt{\spose{\raise 1.5ex\hbox{\hskip3pt$\mathchar"201$}}}    
\def\dt{\spose{\raise 1.0ex\hbox{\hskip2pt$\mathchar"201$}}}    

\def\dotsfill{\leaders\hbox to 1em{\hss.\hss}\hfill}
\def\sun{\odot}

\def\Gyr{{\rm\,Gyr}}
\def\FeH{{\rm[Fe/H]}}

\shorttitle{The PAndAS Field of Streams}
\shortauthors{N. F. Martin et al.}

\begin{document}


\title{The PAndAS Field of Streams: stellar structures in the Milky Way halo toward Andromeda and Triangulum}


\author{Nicolas F. Martin$^{1,2}$, Rodrigo A. Ibata$^1$, R. Michael Rich$^3$, Michelle L. M. Collins$^2$, Mark A. Fardal$^4$, Michael J. Irwin$^5$, Geraint F. Lewis$^6$, Alan W. McConnachie$^7$, Arif Babul$^8$, Nicholas F. Bate$^6$, Scott C. Chapman$^9$, Anthony R. Conn$^6$, Denija Crnojevi\'c$^{10,11}$, Annette M. N. Ferguson$^{10}$, A. Dougal Mackey$^{12}$, Julio F. Navarro$^8$, Jorge Pe\~narrubia$^{10}$, Nial T. Tanvir$^{13}$, David Valls-Gabaud$^{14}$}
\email{nicolas.martin@astro.unistra.fr}

\altaffiltext{1}{Observatoire astronomique de Strasbourg, Universit\'e de Strasbourg, CNRS, UMR 7550, 11 rue de l'Universit\'e, F-67000 Strasbourg, France}
\altaffiltext{2}{Max-Planck-Institut f\"ur Astronomie, K\"onigstuhl 17, D-69117 Heidelberg, Germany}
\altaffiltext{3}{Department of Physics and Astronomy, University of California, Los Angeles, PAB, 430 Portola Plaza, Los Angeles, CA 90095-1547, USA}
\altaffiltext{4}{Department of Astronomy, University of Massachusetts, Amherst, MA 01003, USA}
\altaffiltext{5}{Institute of Astronomy, University of Cambridge, Madingley Road, Cambridge CB3 0HA}
\altaffiltext{6}{Institute of Astronomy, School of Physics A28, University of Sydney, NSW 2006, Australia}
\altaffiltext{7}{NRC Herzberg Institute of Astrophysics, 5071 West Saanich Road, Victoria, BC, V9E 2E7, Canada}
\altaffiltext{8}{Department of Physics and Astronomy, University of Victoria, 3800 Finnerty Road, Victoria, British Columbia V8P 5C2, Canada}
\altaffiltext{9}{Department of Physics and Atmospheric Science, Dalhousie University, 6310 Coburg Road, Halifax NS B3H 4R2, Canada}
\altaffiltext{10}{Institute for Astronomy, University of Edinburgh, Royal Observatory, Blackford Hill, Edinburgh EH9 3HJ, UK}
\altaffiltext{11}{Physics Department, Texas Tech University, Lubbock, TX 79409, USA}
\altaffiltext{12}{RSAA, The Australian National University, Mount Stromlo Observatory, Cotter Road, Weston Creek ACT 2611, Australia}
\altaffiltext{13}{Department of Physics and Astronomy, University of Leicester, University Road, Leicester LE1 7RH, UK}
\altaffiltext{14}{Observatoire de Paris, LERMA, 61 Avenue de l'Observatoire, F-75014 Paris, France}

\begin{abstract}
We reveal the highly structured nature of the Milky Way stellar halo within the footprint of the PAndAS photometric survey from blue main sequence and main sequence turn-off stars. We map no fewer than five stellar structures within a heliocentric range of $\sim5$ to 30 \kpc. Some of these are known (the Monoceros Ring, the Pisces/Triangulum globular cluster stream), but we also uncover three well-defined stellar structures that could be, at least partly, responsible for the so-called Triangulum/Andromeda and Triangulum/Andromeda~2 features. In particular, we trace a new faint stellar stream located at a heliocentric distance of $\sim17\kpc$. With a surface brightness of $\Sigma_V\sim32-32.5 \textrm{ mag}/\textrm{arcsec}^2$, it follows an orbit that is almost parallel to the Galactic plane north of M31 and has so far eluded surveys of the Milky Way halo as these tend to steer away from regions dominated by the Galactic disk. Investigating our follow-up spectroscopic observations of PAndAS, we serendipitously uncover a radial velocity signature from stars that have colors and magnitudes compatible with the stream. From the velocity of eight likely member stars, we show that this stellar structure is dynamically cold, with an unresolved velocity dispersion that is lower than $7.1\kms$ at the 90-percent confidence level. Along with the width of the stream (300--$650\pc$), its dynamics points to a dwarf-galaxy-accretion origin. The numerous stellar structures we can map in the Milky Way stellar halo between 5 and $30\kpc$ and their varying morphology is a testament to the complex nature of the stellar halo at these intermediate distances.
\end{abstract}


\section{Introduction}
The mapping of the structured nature of the stellar halo of the Milky Way (MW) is one of the greatest successes of the Sloan Digital Sky Survey (SDSS). In their `Field of Streams' of the northern Galactic cap, \citet{belokurov06b} highlighted at least four distinct stellar structures that are mainly thought to be the result of the accretion of dwarf galaxies onto the Milky Way over the last few billion years \citep{bell08}. This process of hierarchical formation is a tenet of the currently favored model for the formation of galactic outskirts and is also observed in other nearby galaxies that we can study with enough detail \citep[e.g.][]{ferguson02,martinez-delgado08,martinez-delgado10,mouhcine10}. For instance, the Pan-Andromeda Archaelogical Survey (PAndAS; \citealt{mcconnachie09}) covers a large fraction of the stellar halo of the Andromeda galaxy (M31), and also reveals a wealth of stellar features produced by the tidal disruption of dwarf galaxies \citep{ibata14}.

Although it requires covering large swaths of the sky, imaging the outskirts of the Milky Way remains beneficial to find and track the faintest of stellar streams that cannot be discovered beyond a few tens of kiloparsecs. Discovering these, or proving their absence, is essential if we want to robustly compare the properties of modelled stellar halos with reality \citep{johnston08,cooper10}. Further analyses of the SDSS have illustrated that the Field of Streams is not the whole story, with the discovery of additional thin and/or faint streams from main sequence (MS) and main sequence turn-off (MSTO) stars \citep[e.g.][]{grillmair06,grillmair09,grillmair11,bonaca12,martinc13}. Yet the depth limits of the SDSS and similar surveys like Pan-STARRS~1 are quickly reached in such studies. Until the era of the Large Synoptic Survey Telescope (LSST), almost a decade from now, targeted but deeper surveys can be advantageous for the mapping of the MW stellar halo \citep[e.g.][]{robin07,pila-diez14}, although interpretation can be difficult if angular coverage is limited.

Of particular interest is the region towards M31, which is a target of interest for studies of our cosmic neighbor. These surveys have yielded detections of the Monoceros Ring \citep[MRi;][]{ibata03}, a stellar structure at a heliocentric distance of $\sim7\kpc$ that appears to be circling the disk in the second and third Galactic quadrants and whose nature is a matter of debate (see \citealt{conn12}, and references therein). Beyond the MRi, \citet{majewski04b} discovered a fainter and farther stellar feature, confirmed via the spectroscopy of 2MASS red giant branch stars \citep{rocha-pinto04}. This structure, Triangulum-Andromeda (TriAnd), is estimated to be located at a heliocentric distance of 16 to $25\kpc$, and shows a cloud-like morphology that covers hundreds of square-degrees. This was later confirmed via the analysis of the MW stellar halo in the CFHT pilot program that would become PAndAS \citep{martin07b}. At that time, the imaging data, which only covered a contiguous $\sim76\textrm{ deg}^2$ ($\sim20$\%) of the final survey footprint, showed no spatially well-defined stellar structure in the Milky Way halo, but allowed for the detection of yet another structure, Triangulum-Andromeda 2 (TriAnd2), as an even more distant MS feature in the color-magnitude diagram (CMD). Neither TriAnd, nor TriAnd2 located at a heliocentric distance of $\sim28\kpc$, showed strong spatial variations within the coverage of the survey.

In this paper, we revisit our mapping of the stellar halo of the MW in the region toward M31. We use a large fraction of the full PAndAS coverage ($\sim360\textrm{ deg}^2$) to trace the structure of halo MS and MSTO stars. In Section~2, we briefly describe the PAndAS data we use, while Section~3 presents our mapping of the MW halo and the global properties of the structures. Section~4 focusses on a well-defined stellar stream for which we highlight a possible progenitor and a serendipitous velocity measurement. Finally, we conclude in Section~5.

Where necessary, the distance to the Galactic center is assumed to be $8\kpc$.

\section{Data}
PAndAS is a systematic photometric survey of the surroundings of the Andromeda and Triangulum galaxies. Conducted as a Large Program of the Canada-France-Hawaii Telescope over the period 2008-2011, the survey builds on two original PI programs \citep{ibata07,mcconnachie08} and covers $\sim390\textrm{ deg}^2$ in the $g$ and $i$ bands. The reader interested in the details of the survey, its data reduction, and data products are referred to \citet{ibata14} and A. McConnachie et al. (in preparation), which more thoroughly describe the data set. In this paper, all magnitudes are dereddened following equations~(1) and (2) of \citet{martin13b}.

The regions around M31, M33, NGC 147, and NGC 185 are contaminated by young stars and suffer from crowded photometry, which both tend to populate the region of the CMD dominated by Milky Way halo stellar structures. Therefore, we carve out small areas around these four galaxies, following their overall shape; this lowers the total coverage to $\sim360\textrm{ deg}^2$.

\section{Stellar structures in the direction of Triangulum and Andromeda}
\subsection{The PAndAS CMD}
\begin{figure}
\begin{center}
\includegraphics[width=0.825\hsize,angle=270]{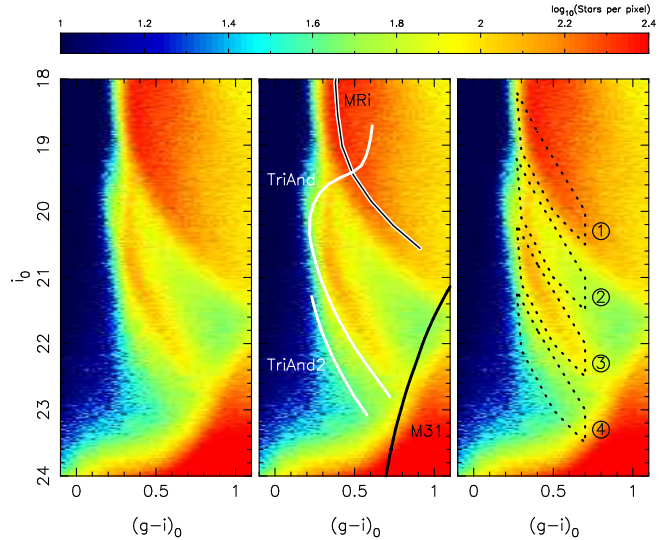}
\caption{\label{CMD_total} The CMD of all stars within PAndAS, excluding masked regions near M31, M33, NGC 147, and NGC 185. Pixels have a size of 0.02~magnitudes in both the color and magnitude directions. Stellar structures at different distances in the Milky Way halo are responsible for the hook-like features visible within the color range $0.2\simlt (g-i)_0 \simlt 1.0$. Stars from the M31 stellar halo produce the dense structures in the faint and red parts of the CMD. The same binned CMD is shown on the three panels, but the MS and MSTO tracks of the features mentioned in the text, along with the region dominated by M31 stars, are overlaid on the middle panel. These are directly taken from \citet{martin07b}. The MS-tailored selection boxes used to build the maps of Figure~\ref{maps} are overlaid in the right-most panel.}
\end{center}
\end{figure}

The CMD of all sources classified as stars in PAndAS is presented in Figure~\ref{CMD_total}, with a focus on the blue colors that correspond to the location of halo main-sequence-turn-off (MSTO) and main-sequence stars (MS). The amount of substructure in this part of the CMD highlights the complexity of the Milky Way stellar halo toward Andromeda and Triangulum, as revealed by earlier contributions \citep[e.g.][]{ibata03,majewski04b,martin07b}. The middle panel shows the same CMD on top of which have been overlaid the tracks of MRi, TriAnd, and TriAnd2 as defined in \citet{martin07b}. The thick disk and the Monoceros Ring create the dense swath of bright stars more luminous than the sharp, curved edge between $((g-i)_0,i_0)=(0.3,19.0)$ and $(1.0,21.2)$. A gap can then be seen before a well-defined MS appears $\sim2$ magnitudes fainter. It is reminiscent of the TriAnd MS discovered by \citet{majewski04b}. In the magnitude range $19.5<i_0<20.5$, this feature extends into a bluer MSTO ($(g-i)_0\simeq0.15$) than that of the MRi as brighter magnitudes ($(g-i)_0\simeq0.30$), thereby highlighting differences in the stellar populations that produce these CMD structures at different heliocentric distances. Further down the CMD, it is not until $i_0\simeq23.0$ that the hook-like MS reaches its lowest density point, just before the contamination from M31's metal-poor red giant branch stars. Below this MS, the faint contribution of the TriAnd2 MS is barely visible in the global PAndAS CMD.

\subsection{Density maps}

In order to map the structure of the MW stellar halo in the PAndAS field, we tailor a selection box to the most prominent MS of Figure~\ref{CMD_total} (box~3 on the right-most panel). Without knowing the properties of the stars that constitute this feature (which, we will see below, is quite complex), it is not possible to accurately determine the mean heliocentric distance this box corresponds to. However, a broad comparison with old and metal-poor isochrones (10 Gyr, $\FeH=-1.5$; see Section~\ref{stellar_pops} below) suggests a mean heliocentric distance of $\sim17\kpc$ for a heliocentric distance range of $14\simlt D_\sun\simlt20\kpc$, indeed compatible with TriAnd \citep{majewski04b,martin07b}. We then proceed to shift this selection box up and down in the CMD in order to map the Monoceros Ring MS stars (box~1, $\sim7\kpc$; $5\simlt D_\sun\simlt8\kpc$), the stars between MRi and TriAnd (box~2, $\sim11\kpc$; $9\simlt D_\sun\simlt13\kpc$), and the stars of TriAnd2, fainter than TriAnd (box~4, $\sim27.5\kpc$; $22\simlt D_\sun\simlt32\kpc$). The spatial homogeneity of the completeness does not vary significantly over the survey for stars in these boxes as these always probe brighter magnitudes than $i_0=23.5$. It should however be noted that the faintest box grazes the metal-poor red giant branch stars of Andromeda and therefore contains some contamination from M31's dwarf galaxies. Since this paper focuses on large-scale structures and the M31-dwarf-galaxy contamination is very localized, this is not an issue. A mild contamination by M31 metal-poor halo stars is also possible for this selection box. Such a contamination would however produce an almost isotropic distribution centered on M31 (Figure~9d of \citealt{ibata14}), which is unlike the stellar features we track from these stars. We are therefore led to conclude that the contamination by M31 halo stars is not an issue for the current analysis.

\begin{figure*}
\begin{center}
\includegraphics[width=1.1\hsize,angle=270]{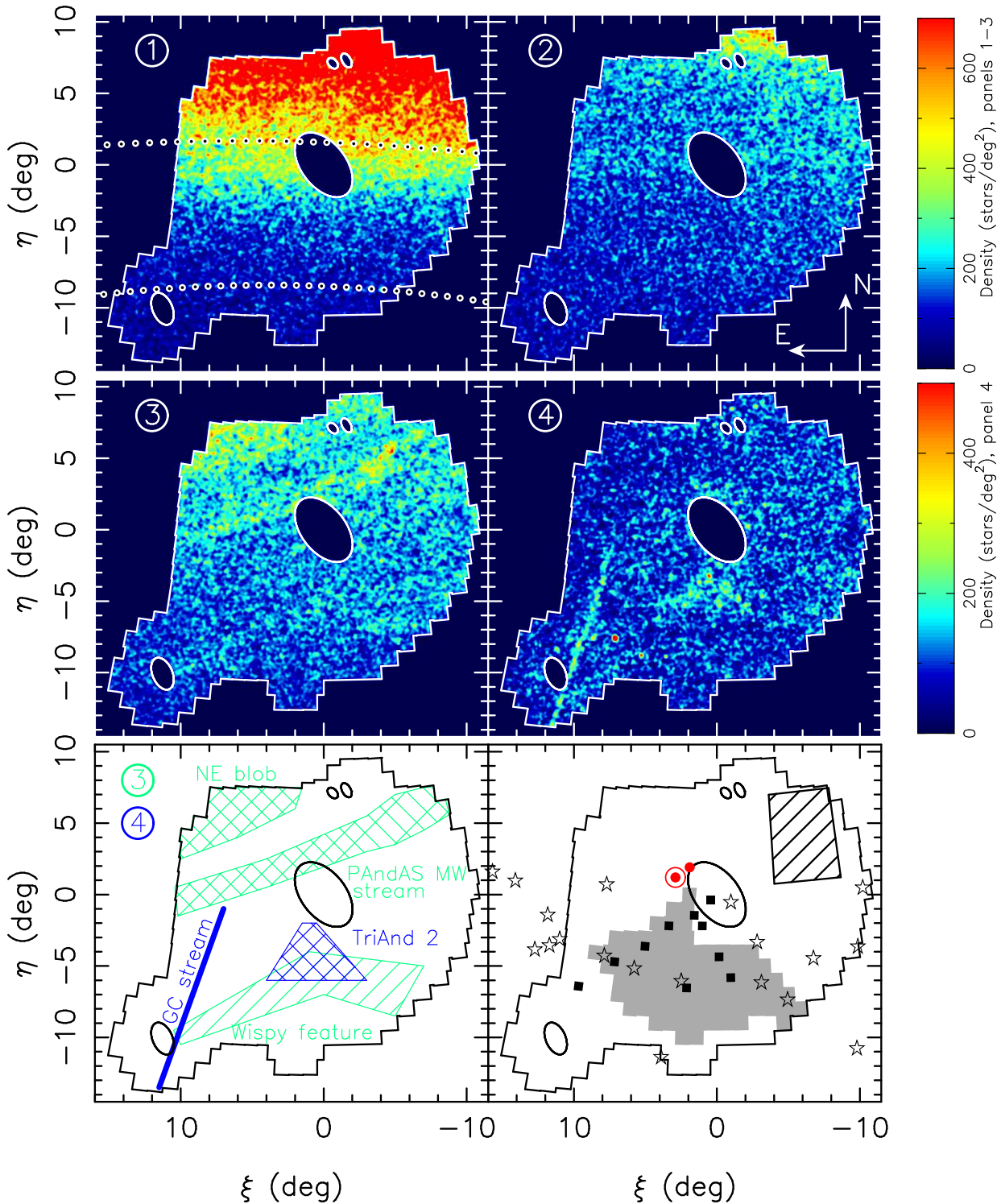}
\caption{\label{maps} \emph{Top two rows of panels:} Smoothed density maps of stars from the four selection boxes highlighted in the right panel of Figure~\ref{CMD_total}. Stars are binned with pixels of $2.5'\times2.5'$ and smoothed with a Gaussian kernel of dispersion $5'$. North is to the top and east to the left, while the color codes the stellar density as indicated in the color scales next to the panels. The white polygon delimits the footprint of the PAndAS survey and the ellipses represent the four regions around M31, M33, NGC 147, and NGC 185 that are masked out to avoid contamination. In the top-left panel, the dotted, curved lines are lines of constant Galactic latitude at $b=-20\deg$ (top) and $b=-30\deg$ (bottom). The panels correspond to heliocentric distances of about 7, 11, 17, and $27\kpc$ from top-left to bottom-right. \emph{Bottom-left panel:} Sketch of the stellar structures visible in panels 3 (green) and 4 (blue) and mentioned in the text. \emph{Bottom-right panel:} Coverage of previous surveys used to discover or study the stellar halo in the direction of M31. The black squares correspond to the fields of \citet{majewski04b} while the gray polygon represents the footprint of the CFHT data that was studied by \citet{martin07b}. The hollow stars show the location of the radial-velocity-selected TriAnd giant stars from \citet{rocha-pinto04}. The region we use in Section~\ref{progenitor} to determine the structure of a potential progenitor to the PAndAS MW stream corresponds to the hashed polygon. The red, circled dot highlights the DEIMOS field in which we find a velocity signal likely stemming from the PAndAS MW stream (see Section~\ref{vels}) and the red dots corresponds to the location of the And~IX dwarf galaxy for which the radial velocities of \citet{tollerud12} show a similar contamination.}
\end{center}
\end{figure*}

The density maps resulting from these four selection boxes are displayed in the top two rows of Figure~\ref{maps} for pixels of $2.5'\times2.5'$, smoothed by a Gaussian kernel of $5'$ dispersion. They reveal varying types and amounts of substructures. To guide the reader, the bottom-left panel of the figure shows a sketch locating the stellar structures we mention in the text and the bottom-right panel locates previous surveys of stellar features in this region.

\emph{Panel 1 ($D_\sun\sim 7\kpc$, $D_\mathrm{GC}\sim13\kpc$)---} This map is expected to trace the density of thick disk and/or MRi stars. \citet{ibata03} showed that the density of these stars changes drastically between the north and south of M31 and we confirm this here from the full PAndAS data set. It is however hard to evaluate how much this change is linked to the thick disk density increasing towards the Galactic plane, which happens to almost coincide with the equatorial north direction over the PAndAS footprint. A generic Galactic model, such as the one of \citet{dejong10b}, reasonably captures the changes of density that we map, provided the thick-disk-to-halo density ratio of the model is increased by a factor of $\sim6$; this is a drastic change to the model and should alert us against such a simplistic comparison model. The presence of density waves in the disk, excited, for instance, by the close passage of a massive dark matter halo \citep{kazantzidis08,purcell11,widrow12,gomez13} or by the disk's spiral arms \citep{siebert12,faure14}, could however displace the disk stars and enhance their contribution to the map of Panel 1 beyond what is assumed in the \citet{dejong10b} model. Eventually, a more global mapping of this stellar structure, such as the one provided by the Pan-STARRS1 survey (C. Slater et al., in preparation), is necessary to deconstruct it.

\emph{Panel 2 ($D_\sun\sim 11\kpc$, $D_\mathrm{GC}\sim17\kpc$)---} The stars of selection box 2, which barely grazes the MRi stars, does not display a significant amount of substructure. The density notably increases from south to north, and its behavior is reminiscent of the MRi map, albeit at much lower densities. The dense region in the north corresponds to regions of increasing Galactic extinction, and could stem from MRi stars bleeding into our selection box from de-reddenning errors.

\emph{Panel 3 ($D_\sun\sim 17\kpc$, $D_\mathrm{GC}\sim22\kpc$)---} This panel traces the very prominent MS that is visible in Figure~\ref{CMD_total} and corresponds to stars at the distance of the TriAnd stellar structure. Contrary to previous attempts at mapping the feature, which revealed no evident spatial structure, thereby hinting at a distant cloud of stars with a significant width \citep{majewski04b,rocha-pinto04,martin07b}, the new extent of PAndAS north of M31 reveals that the stellar halo of the Milky Way is in fact very structured at these distances, $\sim22\kpc$ from the Galactic center. At least two different features are visible in this map: a broader overdense ` NE blob' in the north-east corner of the footprint, which very likely extends further than the limits of the survey, and a stream-like overdensity that crosses the PAndAS footprint, from north-west to east, with an overdense region to the north-west. This stream, which we dub the PAndAS MW stream, will be described in more detail in Section~\ref{stream}.

In addition, we should not forget that TriAnd was initially found south of M31, so it must still be present there at a lower density and, indeed, one can notice a small amount of low-level substructure in the maps of panel 3. In particular, a wispy stream-like feature seems present at the limit of detection, almost parallel to the prominent stream, to its south, running from the masked region around M33 towards the north-west.

\emph{Panel 4 ($D_\sun\sim 27\kpc$, $D_\mathrm{GC}\sim32\kpc$)---} The thin diagonal stream that grazes M33 corresponds to the Pisces/Triangulum stream independently discovered in the SDSS data by \citet{bonaca12} and \citet{martinc13}. This likely globular cluster stream clearly extends well into the Andromeda constellation and abruptly ends $\sim13\deg$ after it enters the PAndAS footprint, further north than it has been traced with the SDSS. The deep PAndAS data leads to a much clearer detection of the stream than was possible with the SDSS data and its thorough photometric analysis will be presented in M. Fardal et al. (in preparation).

Despite this map probing deeper parts of the CMD, the wispy feature that appears south of the masked Andromeda region, in the vicinity of  $(\xi,\eta)=(0\deg,-5\deg)$,  neither correlates with the prominent stellar structures of the M31 halo, nor with the features of the survey's background galaxy distribution, or the foreground Milky Way dust distribution (see Figure~20, 13, and 3 of \citealt{ibata07}, respectively), giving credence that this is yet another genuine stellar structure\footnote{Although the CMD selection box for this panel includes some M31 metal-poor red giant branch stars, these almost exclusively belong to compact dwarf galaxies and are not responsible for the extended features visible in the map. One example of this contamination can be see as the strong overdensity at $(\xi,\eta)\simeq(+7.0\deg,-7.5\deg)$ produced by stars in the And~II dwarf galaxy.}. In addition, this feature is still present in the maps, albeit at lower significance, when restricting the selection to stars brighter than $i_0=22.8$ that should not be contaminated by M31 stellar populations, or background galaxies. The CMD of this region shows MS stars that have fainter magnitudes than the prominent MS visible in Figure~1. Since it overlaps with the footprint of the data at the time of the analysis of \citet{martin07b}, it likely corresponds to the TriAnd2 structure discovered in that work. The latest PAndAS photometric calibration \citep{ibata14}, combined with the re-observation of some low-quality fields that covered most of the region of this overdensity, means that it is now possible to trace the extent of this cloud of stars\footnote{It is worth noting that, although the map of the TriAnd2 stars built in \citet[][the right-most panel of their Figure~3]{martin07b} showed no well-defined overdensity, regions that overlap with the wispy feature we identify here did show a mild density increase.}. It roughly covers a region of $\sim3\deg\times5\deg$ which, at this distance, corresponds to a physical extent of $\sim1.5\times2.5\kpc^2$.

\begin{figure*}
\begin{center}
\includegraphics[width=\hsize]{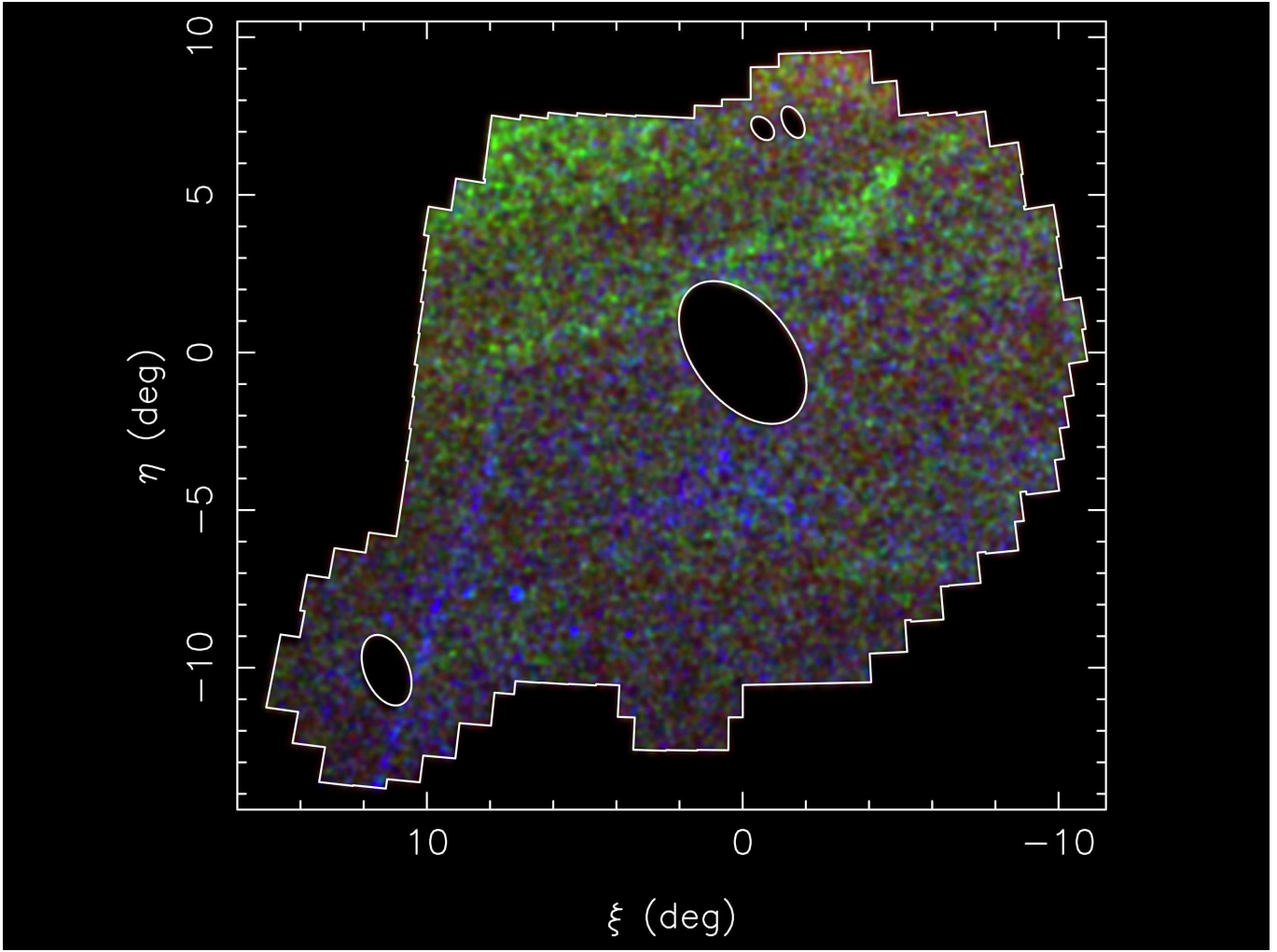}
\caption{\label{map_RGB} The PAndAS `Field of Streams' built from panels 2 (red; $D_\mathrm{GC}\sim17\kpc$), 3 (green; $D_\mathrm{GC}\sim22\kpc$), and 4 (blue; $D_\mathrm{GC}\sim32\kpc$) of Figure~\ref{maps}, displaying the highly structured nature of the Milky Way halo in the direction of Andromeda and Triangulum. North is to the top, and east to the left.}
\end{center}
\end{figure*}

Figure~\ref{map_RGB} combines panels 2, 3, and 4, to produce a color image of the stellar halo between roughly 10 and $30\kpc$. This `PAndAS Field of Streams' stresses the highly structured nature of the MW stellar halo in the survey's cone towards M31 with a crisscrossing of stellar features of varying morphology, density, and distances. It is worth remembering at this stage that the PAndAS footprint covers less than 1 percent of the sky. It is therefore to be expected that, despite the revolution provided by panoramic sky surveys, many halo stellar structures have so far gone unnoticed due to their limited coverage and/or depth.

\subsection{Stellar populations}
\label{stellar_pops}

\begin{figure*}
\begin{center}
\includegraphics[width=0.32\hsize,angle=270]{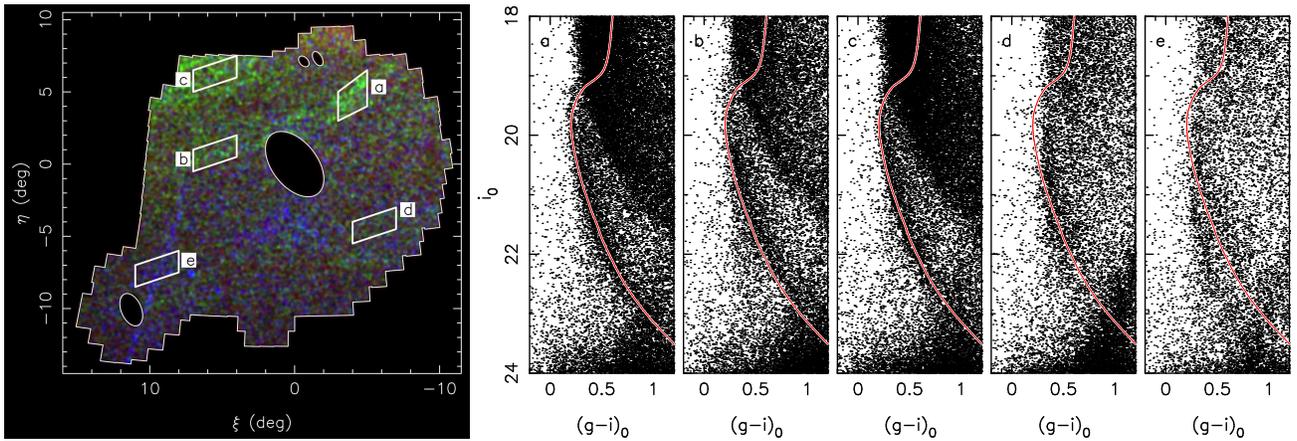}
\includegraphics[width=0.32\hsize,angle=270]{f4b.ps}
\caption{\label{CMDs} Color-magnitude diagrams of five 4.5 deg$^2$ fields in the PAndAS footprint. The locations of the fields are highlighted in the map displayed in the left-most panel, which reproduces panel 3 of Figure~\ref{maps}. All five fields have the same coverage and target regions of interest, as described in the text (Section~\ref{stellar_pops}). A Dartmouth isochrone of age 10~Gyr, metallicity $\FeH=-1.5$ \citep{dotter07}, and $[\alpha/\textrm{Fe}]=+0.2$, shifted to a distance modulus of 16.2 (i.e. a distance of $17\kpc$), is overlaid on the CMDs and shows a good match with the MS and MSTO of fields (a) and (c).}
\end{center}
\end{figure*}

In Figure~\ref{CMDs}, we display the CMDs of five different fields which target various regions of interest within PAndAS (the fields are marked on the map in the left-most panel of the figure). Each field has been tailored to have the same spatial coverage for an easier comparison of the MS features they display. Field (a) focusses on the overdensity in the new stream, whereas field (b) targets the stream itself. Their CMDs are very similar in both cases, displaying a very well defined MS that is suitably represented by a Dartmouth isochrone \citep{dotter07} of $10\Gyr$, with a metallicity $\FeH=-1.5$ and $[\alpha/\textrm{Fe}]=+0.2$, shifted to a distance modulus of 16.2, or $17\kpc$. Such an isochrone nicely follows the curvature of the MS, and correctly reaches the MSTO of this stellar population that is significantly bluer than the MSTO of the thick disk and MRi ($(g-i)_0\sim0.15$ for the isochrone vs. $(g-i)_0\sim0.30$ for the bluest MRi stars). It should be noted, however, that changing the isochrone parameters to $13\Gyr$ and $\FeH=-2.0$ yields an equally good match. Using the isochrone as a guide, one can note that the MS in panel (b) is slightly closer/brighter than that of panel (a), pointing towards a distance gradient along the piece of the stream that intersects the PAndAS footprint. A shift of the isochrone to $\sim0.3$ magnitudes brighter provides a better match to the MS of panel (b). This is a significant distance gradient since it means that the stream is getting $\sim2\kpc$ closer to us over the $\sim3\kpc$ that separate the two fields.

The CMD of the field in the `NE blob' (panel (c)) displays a MS that is very similar to the one in panel (a) and the same isochrone as above is also a good match for a distance modulus of 16.2. Since there is no distinct bridge of stars between the stream and the `blob' and it is hard to imagine a scenario where the stream, which has to orbit the MW, sharply curves beyond the edge of the PAndAS footprint to form the `blob,' we are left to conclude that the MW halo hosts two unrelated but similar stellar structures at a similar distance, very close-by on the sky. Spectroscopic follow-up is needed to verify this scenario since it would be unlikely for two unrelated halo structures to fill neighboring parts of phase-space.

Even fields that track no obvious features, such as fields (d) and (e), still have CMDs that display a MS, even though it is much more diffuse than in the case of the well-defined MS of panels (a)--(c). This is the TriAnd MS discovered by \citet{majewski04b} and that we tracked in the pre-PAndAS CFHT data \citep{martin07b}. A comparison of this MS with the one that appears in the first three panels makes it evident that the TriAnd MS is broader and more poorly defined. This is likely a consequence of TriAnd having a larger extent along the line of sight than the stream and the `blob.' This was already suggested by the 3D maps of the structure obtained from 2MASS giant stars by \citet{rocha-pinto04}, and is compatible with the large extent of TriAnd on the sky as well as its cloud-like morphology. The redder MSTO of TriAnd suggests that its stars are more metal-rich and/or older than those of the stream and `blob.' The former is compatible with the spectroscopic measurements of \citet{rocha-pinto04}, which yielded an average metallicity of $\FeH=-1.2$ for TriAnd red giant branch stars.

Finally, panel (e) displays the additional MS of the Pisces/Triangulum globular cluster stream at fainter magnitudes.

\section{The PAndAS MW stream}
\label{stream}
The most prominent new feature of the MW stellar halo revealed by the PAndAS data is the east-west stream visible in panel 3 of Figure~\ref{maps}. We have seen above that, at the distance to the stream, we intersect more than $3\kpc$ of its extent, but that it shows a significant distance gradient of $\sim2\kpc$ over this length. The stream also has a measurable angle with respect to the Milky Way plane as, between fields (a) and (b) highlighted on Figure~\ref{CMDs}, its height below the plane increases from $\sim4.7\kpc$ to $\sim5.4\kpc$, crudely leading to a pitch angle of slightly more than $15\deg$. Of course, a larger area is required to properly characterize the detailed shape and morphology of the stream.

\subsection{Physical dimensions}
\begin{figure}
\begin{center}
\includegraphics[width=0.7\hsize,angle=270]{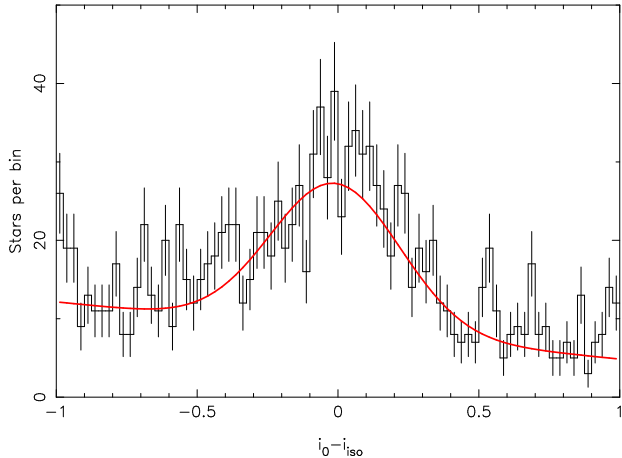}
\caption{\label{mag_profile} Distribution of magnitude distances from the red isochrone of Figure~\ref{CMDs} for stream stars in field (a) and within the color range $0.5<(g-i)_0<0.7$. The error bars on the histogram counts correspond to Poissonian uncertainties. The red line corresponds to the best fit model of the sum of a slope to represent the contamination and a Gaussian to represent the stream stars. The Gaussian has a width $\delta\mu = 0.24$~magnitudes.}
\end{center}
\end{figure}

The extent of the stream on the sky is $\sim$1--2$\deg$, which, assuming a distance of $\sim17\kpc$, translates to a physical width of 300--650\pc. Measuring the size of the stream perpendicular to the line of sight is obviously a more uncertain endeavor, but the narrowness of its MS feature in the CMD of Figure~\ref{CMDs} can nevertheless yield an upper limit. Following the method presented in \citet{sollima11}, we construct the profile of magnitude distances from the red isochrone of Figure~\ref{CMDs} for all stars with $0.50<(g-i)_0<0.70$. The resulting histogram, displayed in Figure~\ref{mag_profile}, shows the Gaussian overdensity of the stream's MS stars without ambiguity, centered on the isochrone ($i_0-i_\mathrm{iso}\simeq0.0$). Under the assumption that the stream is composed of a single stellar population, the width of this Gaussian is entirely due to the line of sight extent of the structure. Fitting this profile with a simple model that is the sum of a background slope and a Gaussian (the red line overlaid on the histogram) yields the width of the overdensity: $\delta\mu = 0.23$~magnitudes. From this quantity, it is straightforward to calculate the line-of-sight distance of the stream as $\delta D = 0.461\delta\mu D = 0.11D$. Once again assuming a heliocentric distance of $\sim17\kpc$ to the structure, we find its depth along the line of sight is less than 1.8\kpc.

Of course, the width of the Gaussian in Figure~\ref{mag_profile} is unlikely to be entirely due to scatter in the physical distance to the stream stars. If the photometric uncertainties at these magnitudes remain quite small ($\lta0.05$~magnitudes) and likely don't inflate the profile, the presence of metallicity and/or age spreads in the stream's stellar populations would certainly lead to us overestimating the true light-of-sight extent of the PAndAS MW stream. It is nevertheless clear that this stream is confined in space and does not correspond to a very extended cloud-like structure like the bulk of TriAnd \citep{rocha-pinto04}.

\subsection{Surface brightness}
Estimating the surface brightness of this stellar feature is rendered difficult by the complex nature of the stellar halo, and the difficulty in disentangling this stream from other structures. However, using $10\Gyr$ and $\FeH=-1.5$ \textsc{Parsec} isochrones and luminosity functions \citep{bressan12}, we nevertheless construct a color-magnitude probability density function (pdf) of the stellar population of the stream, convolved by the observational uncertainties, that we then populate until we reach the relevant number of stars in the color-magnitude selection box 3 shown in Figure~\ref{CMD_total} for fields (a) and (b). Adding the contribution of all drawn stars and normalizing it to the coverage of the fields yields the average surface brightness of the stream. We choose two bracketing alternatives to calculate the number of stars in the CMD box. The high estimate assumes that all stars are stream-members (an unlikely case since the stellar halo north and south of the stream shows non-negligible counts in Figure~\ref{maps}) and yields maximal $V$-band surface brightnesses of $\sim31\textrm{ mag}/\textrm{arcsec}^2$ for both fields (a) and (b), respectively. Our low estimate is calculated by subtracting the counts of the contamination model developed by \citet{martin13b}. This model fits an exponential density function in $\xi$ and $\eta$ for any location of the CMD, but has the drawback of incorporating parts of the stream in the data used to determine the preferred density functions. This leads to an oversubtraction of the counts in the CMD selection box, and low estimates of the V-band surface brightness of $32.5\textrm{ mag}/\textrm{arcsec}^2$ and $32.7\textrm{ mag}/\textrm{arcsec}^2$, for fields (a) and (b), respectively. The true surface brightness of these two locations of the stream are therefore likely between $32.0$ and $32.5\textrm{ mag}/\textrm{arcsec}^2$, comparable to previous estimates for the TriAnd and TriAnd2 stellar structures \citep{majewski04b,martin07b}.

\subsection{A possible progenitor?}
\label{progenitor}
\begin{figure*}
\begin{center}
\includegraphics[width=0.4\hsize,angle=270]{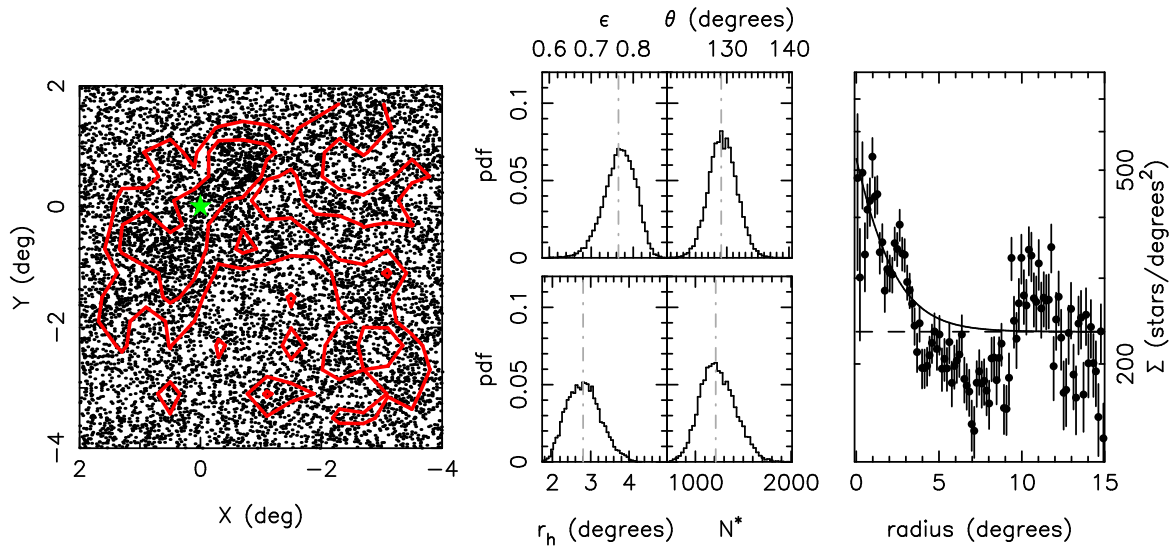}
\caption{\label{stream_overdensity}The structure of the overdensity in the PAndAS MW stream. The left-most panel displays the spatial distribution of all stars in the CMD selection box 3 of Figure~\ref{CMD_total} at the location of the overdensity. The red contours are isodensity contours of 250 and 345 stars per square degree, built from a smoothed density map. While the overdensity is obvious, its shape is irregular. The green star indicates its centroid, as derived from the determination of its structural parameters. The middle set of four panels displays the marginalized probability distribution functions (pdfs) of four parameters of the structural model: the ellipticity, $\epsilon$, defined as $1-b/a$ with $a$ and $b$ the major and minor axis scale length, respectively; the position angle, $\theta$, defined from north to east; the half-light radius of the exponential profile, $r_h$; the number of overdense stars within the CMD selection box, $N^*$. In all four panels, the gray dot-dashed line represents the favored model, i.e. the mode of the marginalized pdfs. The right-most panel presents the profile of the data, binned following the favored model parameters, with the favored exponential profile with a flat background (thin black line). The background alone is represented by the dashed line. The fit is obviously not a good one as it cannot capture the varying structure of the field. However, it reasonably reproduces the central overdensity, albeit with too shallow a profile.}
\end{center}
\end{figure*}

The stellar stream, with a width of 300--650$\pc$ is likely produced by the tidal disruption of a dwarf galaxy rather than a globular cluster onto the Milky Way, following an orbit close to planar with the Milky Way disk. There is no clear evidence of its progenitor over the $\sim20\deg$ (or $\sim6\kpc$) it spans within PAndAS, although there is a marked overdensity within this stream, north-west of M31, that is not unlike the Bo\"otes~III overdensity discovered in the SDSS by \citet{grillmair09}. The distribution of stars selected by the CMD box labeled 3 in Figure~\ref{CMD_total} is displayed in Figure~\ref{stream_overdensity} around this overdensity. The boundaries of this region were tailored so as to avoid most fluctuations of the field density and steer away from the stream's east/west extent. We apply the technique of \citet[][update in N. F. Martin et al., in preparation, to a full Markov-Chain Monte Carlo treatment]{martin08b} to fit the best flattened exponential radial density model to this stellar distribution and it converges, but yields a poor fit, as shown in Figure~\ref{stream_overdensity}. Although the pdfs of parameters that are fitted for are well behaved (middle panels), with a center located at $(\ell,b)=(117.2\deg,-16.6\deg)$, the best profile clearly cannot account for the very structured field density. Nevertheless, the fit gives a rough idea of the size of this overdensity, $r_h\sim2.8\deg$ along its major axis, which corresponds to a physical extent of $r_h\sim830\pc$ at $17\kpc$. With an ellipticity of $\sim0.8$ the structure of this overdensity is driven by stream stars, probably inflating our measurement of $r_h$, but it is also not unlike Bo\"otes~III or Ursa Major~I found farther out in the halo of the MW \citep{correnti09,martin08b}. Spectroscopic follow-up will however be necessary to determine if this overdense part of the stream is still a bound dwarf galaxy progenitor of the stream or, as seems more likely, a clump of stream stars.

\subsection{Velocities}
\label{vels}
\begin{figure}
\begin{center}
\includegraphics[width=0.8\hsize,angle=270]{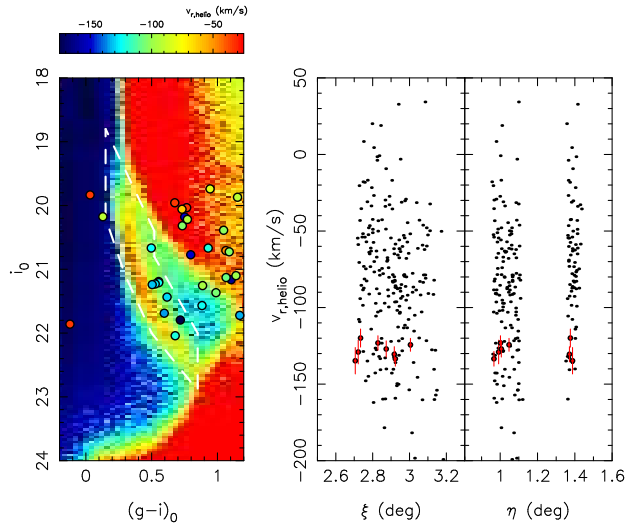}
\caption{\label{velocities}The left-most panel displays the CMD of Figure~\ref{CMD_total} over which we have represented the stars from a series of three DEIMOS fields located at $(\xi,\eta)\sim(+3\deg,+1\deg)$. These fields overlap the PAndAS MW stream and show a radial velocity signature. Stars observed with DEIMOS are color-coded by their heliocentric radial velocities, as shown above the panel. The two panels on the right-hand side show the velocity of all the MW stars in these fields as black dots. The subsample of stars which fall in the white dot-dashed selection box overlaid on the CMD are shown in red. They are all clumped at a well-defined systemic velocity, except for one star with a velocity of $-263\pm5\kms$ that falls beyond the displayed velocity range.}
\end{center}
\end{figure}

With the discovery of the stream, we went back to the library of radial velocities we have gathered over the years with DEIMOS/Keck on the stellar structures of the M31 halo \citep[e.g.][]{ibata05,chapman06,collins13} and searched for a serendipitous velocity signature of the stream among the foreground Milky Way stars. A series of three DEIMOS fields located near $(\xi,\eta)\sim(+3\deg,+1\deg)$, and represented by the circled red dot in the bottom-right panel of Figure~\ref{maps}, overlap with the PAndAS MW stream and show a cold velocity signature. The left panel of Figure~\ref{velocities} displays the PAndAS CMD over which were added all observed stars with reliable radial velocities ($4-8\kms$) from these three fields\footnote{The selection function that led to the observation of these stars is very complex, and focusses on M31 stars, which were the main focus of the spectroscopic program. Therefore, the presence/absence of stars in specific regions of the sky or the CMD is no indication of the presence/absence of CMD features.}; a handful of these stars fall close to the MS of the stream. If one star selected within the dashed white box of Figure~\ref{velocities} has a large negative velocity ($-263\pm5\kms$) and is likely a MW or M31 halo star, the other eight stars within this box clump at a very similar velocity, near $v_\mathrm{r,helio} = -125\kms$ (the red points in the right panels of the figure). If these were random halo or disk stars, they would scatter over a large velocity range like the other stars in the sample, shown in black.

\begin{figure}
\begin{center}
\includegraphics[width=0.7\hsize,angle=270]{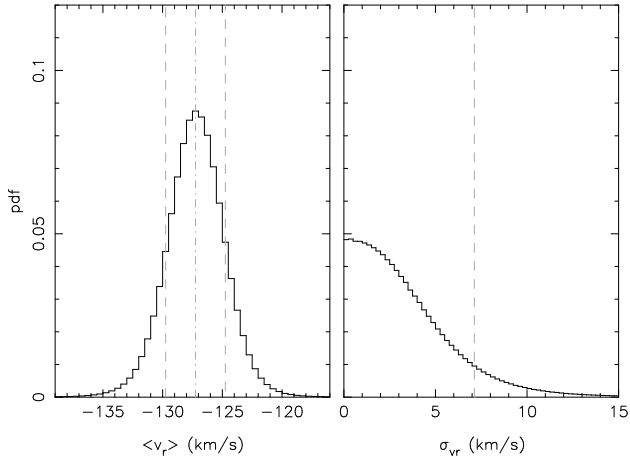}
\caption{\label{velocities_pdfs}The pdfs of the heliocentric systemic velocity, $\langle v_\mathrm{r}\rangle$, and velocity dispersion, $\sigma_\mathrm{vr}$, of the eight stream stars shown in red in the right-most panels of Figure~\ref{velocities} that belong to the PAndAS MW stream. The grey dot-dashed lines highlight the favored parameters and the grey dotted line brackets the $\pm1$-$\sigma$ confidence interval, or indicate the 90-percent confidence upper limit in the case of $\sigma_\mathrm{vr}$.}
\end{center}
\end{figure}

Our sample of stream stars with velocities is small, but can nevertheless be used to constrain the dynamical properties of the stellar stream. We fit the velocity distribution of the stream members with a Gaussian model, following a likelihood technique similar to the one described in \citet{martin07a}.  The pdfs of the two parameters we fit for (the heliocentric systemic velocity, $\langle v_\mathrm{r}\rangle$, and the velocity disperison, $\sigma_\mathrm{vr}$) are displayed in Figure~\ref{velocities_pdfs} for flat priors. If the systemic velocity of the stream at the location of the field is well constrained at $\langle v_\mathrm{r}\rangle = -127.2\pm2.5\kms$, the population is too cold to resolve the velocity dispersion, despite relatively small velocity uncertainties of $\sim4$--$8\kms$. The velocity dispersion is constrained to be lower than $7.1\kms$ at the 90-percent confidence level.

Interestingly, \citet{tollerud12} found a very similar signal in their own DEIMOS observations of the dwarf galaxy And~IX, located $\sim1\deg$ east of our detection (the red dot in the bottom-right panel of Figure~\ref{maps}), for which they measure $v_\mathrm{r,helio} \simeq -130\kms$. Although they unfortunately did not publish the properties of these stars and we therefore cannot study their relation to the prominent MS and MSTO of the PAndAS MW stream, the authors point out that the stars forming this radial velocity peak do not share the color and magnitude expected for old M31 red giant branch stars. It is very likely that both our data and theirs detected the kinematic signature of the same structure.

Could both detections nevertheless be produced by M31 giant stars? At the location of these fields the M31 extended disk, if present, is expected to have velocities between $-100$ and $-200\kms$ \citep{ibata05}, not unlike what we measure here. There are also young stellar populations (0.2--2~Gyr) at quite large distances from M31 \citep{richardson08}, whose colors are bluer than the bulk of M31 red giant branch stars visible in the CMD of Figure~\ref{CMD_total}. At present, it is therefore not possible to entirely rule out that the velocity signal stems from these M31 stars. However, the similar radial velocity measured here and by \citet{tollerud12} in front of And~IX points to a structure whose velocity remains almost constant over $\sim1\deg$. At the distance of M31, this separation translates to $\sim15\kpc$ and the observed heliocentric velocity of disk stars this far apart is expected to change significantly (see, e.g., Figure~7 of \citealt{ibata05}). Of course, we cannot rule out that the stars in radial velocity peak would be stirred up stars from the M31 disk whose kinematics would not faithfully track those of the disk.

Overall, we however deem it more likely that these stars, given their location in the CMD and the lack of changes in their velocity with position, trace the stellar feature we identify crossing panel 3 of Figure~\ref{maps}. Their dynamical properties bolster the scenario of a stellar stream, whose progenitor was disrupted by the MW. A MW-disk-related structure would be dynamically hotter, with a velocity dispersion upwards of $30\kms$ \citep[e.g.][although it should be noted that this measurement stems from closer stars]{kordopatis13}. It remains to be seen whether the progenitor of the stream really is the overdensity mapped above and, if so, whether it is a dwarf galaxy or a globular cluster. However, with a width of 300--650$\pc$, the former appears more likely.

Interestingly, the morphology and the systemic velocity of the PAndAS MW stream has some similarities with the \citet{penarrubia05} simulation of the Monoceros Ring produced by a satellite accretion in the plane of the Milky Way. The simulation particles within a distance range equivalent to that of panel~3 of Figure~\ref{maps} also produce a stream that cuts through the PAndAS footprint along the east-west axis. Although the fine details of the simulated and observed streams differ, it does hint that the feature we discovered in the foreground of PAndAS was brought in by a system on a low-inclination, low-eccentricity orbit, potentially also responsible for closer stellar features such as the MRi.

\section{Conclusions}

From the analysis of blue MS and MSTO stars in the PAndAS footprint, we have shown that the Milky Way halo is very structured in the direction of Andromeda and Triangulum, out to distances of at least $\sim30\kpc$. In addition to the stellar features already known in this part of the sky --- the Monoceros Ring \citep{ibata03}, Triangulum/Andromeda \citep{majewski04b,rocha-pinto04}, Triangulum/Andromeda~2 \citep{martin07b}, the Pisces-Triangulum globular cluster stream \citep{bonaca12,martinc13} --- we reveal the presence of at least two new stellar features. At $D_\sun\sim17\kpc$, a stream-like structure crosses the PAndAS footprint, mainly along the east-west axis that is coplanar with the Galactic disk, and another `blob,' which is cut off by the north-eastern limits of the survey. The stellar populations that compose these two structures are very similar, in both cases old and metal-poor, but there is no obvious connection between them. The stream exhibits a strong radial distance gradient, with its heliocentric distance diminishing by $2\kpc$ over a sky-projected path of only $\sim3\kpc$. It has a width of a few hundred parsecs and a depth smaller than $1.8\kpc$, implying it is likely the product of the disruption of a dwarf galaxy, whose remnant may be a localized overdensity embedded in the stream, located at $(\ell,b)\sim(117.2\deg,-16.6\deg)$. We further find a kinematic signature of likely stream stars in our library of DEIMOS/Keck spectra targeting M31 stellar structure. We show that the PAndAS MW stream star candidates are dynamically cold with a velocity dispersion of no more than $7.1\kms$ at the 90-percent confidence level.

This analysis of the PAndAS foreground reveals that the SDSS `Field of Streams' \citep{belokurov06b} is but the tip of the iceberg. Reaching fainter/more diffuse structures, even over the comparatively small PAndAS area ($\sim360\textrm{deg}^2$), unveils many substructures which would be difficult to map convincingly without the exquisite PAndAS photometry. Yet PAndAS does not even cover 1 percent of the night sky. One can therefore expect that surveys like the DES\footnote{http://www.darkenergysurvey.org}, or the LSST\footnote{http://www.lsst.org/lsst/}, will reveal another wealth of stellar streams within at least $50\kpc$ from the Galactic center. This is not unexpected from simulations of the hierarchical build-up of stellar halos \citep[e.g.][]{johnston08,helmi11}, and neither are the morphological differences between the structures. In the PAndAS Field of Streams, we find stream-like, cloud-like, and `blob'-like features, as one can expect at these distances that transition between the very inner parts of the halo, for which dynamical times are short, and its outer regions, for which these can reach a Hubble time.

Finally, the discovery of the new stream is testament that, even though most streams known to date are on orbits close to polar \citep{pawlowski13a}, more planar streams do exist. Our tally of streams may well be biased away from planar orbits, especially if dynamical friction is efficient at aligning the orbit of non-polar accretions with the MW plane \citep{taylor01,penarrubia02,penarrubia06}, effectively hiding these accretions from most surveys that shy away from the disk-dominated regions.

\acknowledgments
N.F.M. thanks Benoit Famaey for interesting discussions and acknowledges partial support by the DFG through the SFB 881 (sub-project A2). G.F.L thanks the Australian research council for support through his Future Fellowship (FT100100268) and Discovery Project (DP110100678). We are grateful to the CFHT observing team for gathering the PAndAS images, and for their continued support throughout the project. This work is based on observations obtained with MegaPrime/MegaCam, a joint project of CFHT and CEA/DAPNIA, at the Canada-France-Hawaii Telescope, which is operated by the National Research Council (NRC) of Canada, the Institut National des Sciences de l'Univers of the Centre National de la Recherche Scientifique (CNRS) of France, and the University of Hawaii. Some of the data presented herein were obtained at the W.M. Keck Observatory, which is operated as a scientific partnership among the California Institute of Technology, the University of California and the National Aeronautics and Space Administration. The Observatory was made possible by the generous financial support of the W.M. Keck Foundation. The authors wish to recognize and acknowledge the very significant cultural role and reverence that the summit of Mauna Kea has always had within the indigenous Hawaiian community.  We are most fortunate to have the opportunity to conduct observations from this mountain.


\begin{thebibliography}{50}
\expandafter\ifx\csname natexlab\endcsname\relax\def\natexlab#1{#1}\fi

\bibitem[{{Bell} {et~al.}(2008){Bell}, {Zucker}, {Belokurov}, {Sharma},
  {Johnston}, {Bullock}, {Hogg}, {Jahnke}, {de Jong}, {Beers}, {Evans},
  {Grebel}, {Ivezi{\'c}}, {Koposov}, {Rix}, {Schneider}, {Steinmetz}, \&
  {Zolotov}}]{bell08}
{Bell}, E.~F., {Zucker}, D.~B., {Belokurov}, V., et al. 2008, \apj, 680, 295

\bibitem[{{Belokurov} {et~al.}(2006){Belokurov}, {Zucker}, {Evans}, {Gilmore},
  {Vidrih}, {Bramich}, {Newberg}, {Wyse}, {Irwin}, {Fellhauer}, {Hewett},
  {Walton}, {Wilkinson}, {Cole}, {Yanny}, {Rockosi}, {Beers}, {Bell},
  {Brinkmann}, {Ivezi{\'c}}, \& {Lupton}}]{belokurov06b}
{Belokurov}, V., {Zucker}, D.~B., {Evans}, N.~W., et al. 2006, \apjl, 642,
  L137

\bibitem[{{Bonaca} {et~al.}(2012){Bonaca}, {Geha}, \&
  {Kallivayalil}}]{bonaca12}
{Bonaca}, A., {Geha}, M., \& {Kallivayalil}, N. 2012, \apjl, 760, L6

\bibitem[{{Bressan} {et~al.}(2012){Bressan}, {Marigo}, {Girardi}, {Salasnich},
  {Dal Cero}, {Rubele}, \& {Nanni}}]{bressan12}
{Bressan}, A., {Marigo}, P., {Girardi}, L., et al. 2012, \mnras, 427, 127

\bibitem[{{Chapman} {et~al.}(2006){Chapman}, {Ibata}, {Lewis}, {Ferguson},
  {Irwin}, {McConnachie}, \& {Tanvir}}]{chapman06}
{Chapman}, S.~C., {Ibata}, R., {Lewis}, et al. 2006, \apj, 653, 255

\bibitem[{{Collins} {et~al.}(2013){Collins}, {Chapman}, {Rich}, {Ibata},
  {Martin}, {Irwin}, {Bate}, {Lewis}, {Pe{\~n}arrubia}, {Arimoto}, {Casey},
  {Ferguson}, {Koch}, {McConnachie}, \& {Tanvir}}]{collins13}
{Collins}, M.~L.~M., {Chapman}, S.~C., {Rich}, R.~M., et al. 2013, \apj, 768, 172

\bibitem[{{Conn} {et~al.}(2012){Conn}, {No{\"e}l}, {Rix}, {Lane}, {Lewis},
  {Irwin}, {Martin}, {Ibata}, {Dolphin}, \& {Chapman}}]{conn12}
{Conn}, B.~C., {No{\"e}l}, N.~E.~D., {Rix}, H.-W., et al. 2012, \apj, 754, 101

\bibitem[{{Cooper} {et~al.}(2010){Cooper}, {Cole}, {Frenk}, {White}, {Helly},
  {Benson}, {De Lucia}, {Helmi}, {Jenkins}, {Navarro}, {Springel}, \&
  {Wang}}]{cooper10}
{Cooper}, A.~P., {Cole}, S., {Frenk}, C.~S., et al. 2010, \mnras, 406, 744

\bibitem[{{Correnti} {et~al.}(2009){Correnti}, {Bellazzini}, \&
  {Ferraro}}]{correnti09}
{Correnti}, M., {Bellazzini}, M., \& {Ferraro}, F.~R. 2009, \mnras, 397, L26

\bibitem[{{de Jong} {et~al.}(2010){de Jong}, {Yanny}, {Rix}, {Dolphin},
  {Martin}, \& {Beers}}]{dejong10b}
{de Jong}, J.~T.~A., {Yanny}, B., {Rix}, H.-W., et al. 2010, \apj, 714, 663

\bibitem[{{Dotter} {et~al.}(2007){Dotter}, {Chaboyer}, {Jevremovi{\'c}},
  {Baron}, {Ferguson}, {Sarajedini}, \& {Anderson}}]{dotter07}
{Dotter}, A., {Chaboyer}, B., {Jevremovi{\'c}}, D., et al. 2007, \aj, 134, 376

\bibitem[{{Faure} {et~al.}(2014){Faure}, {Siebert}, \& {Famaey}}]{faure14}
{Faure}, C., {Siebert}, A., \& {Famaey}, B. 2014, MNRAS, in press
  (ArXiv:1403.0587)

\bibitem[{{Ferguson} {et~al.}(2002){Ferguson}, {Irwin}, {Ibata}, {Lewis}, \&
  {Tanvir}}]{ferguson02}
{Ferguson}, A.~M.~N., {Irwin}, M.~J., {Ibata}, R.~A., {Lewis}, G.~F., \&
  {Tanvir}, N.~R. 2002, \aj, 124, 1452

\bibitem[{{G{\'o}mez} {et~al.}(2013){G{\'o}mez}, {Minchev}, {O'Shea}, {Beers},
  {Bullock}, \& {Purcell}}]{gomez13}
{G{\'o}mez}, F.~A., {Minchev}, I., {O'Shea}, B.~W., et al. 2013, \mnras, 429, 159

\bibitem[{{Grillmair}(2009)}]{grillmair09}
{Grillmair}, C.~J. 2009, \apj, 693, 1118

\bibitem[{{Grillmair}(2011)}]{grillmair11}
---. 2011, \apj, 738, 98

\bibitem[{{Grillmair} \& {Dionatos}(2006)}]{grillmair06}
{Grillmair}, C.~J., \& {Dionatos}, O. 2006, \apjl, 641, L37

\bibitem[{{Helmi} {et~al.}(2011){Helmi}, {Cooper}, {White}, {Cole}, {Frenk}, \&
  {Navarro}}]{helmi11}
{Helmi}, A., {Cooper}, A.~P., {White}, S.~D.~M., et al. 2011, \apjl, 733, L7

\bibitem[{{Ibata} {et~al.}(2005){Ibata}, {Chapman}, {Ferguson}, {Lewis},
  {Irwin}, \& {Tanvir}}]{ibata05}
{Ibata}, R., {Chapman}, S., {Ferguson}, A.~M.~N., et al. 2005, \apj, 634, 287

\bibitem[{{Ibata} {et~al.}(2007){Ibata}, {Martin}, {Irwin}, {Chapman},
  {Ferguson}, {Lewis}, \& {McConnachie}}]{ibata07}
{Ibata}, R., {Martin}, N.~F., {Irwin}, M., et al. 2007, \apj, 671, 1591

\bibitem[{{Ibata} {et~al.}(2003){Ibata}, {Irwin}, {Lewis}, {Ferguson}, \&
  {Tanvir}}]{ibata03}
{Ibata}, R.~A., {Irwin}, M.~J., {Lewis}, G.~F., {Ferguson}, A.~M.~N., \&
  {Tanvir}, N. 2003, \mnras, 340, L21

\bibitem[{{Ibata} {et~al.}(2014){Ibata}, {Lewis}, {McConnachie}, {Martin},
  {Irwin}, {Ferguson}, {Babul}, {Bernard}, {Chapman}, {Collins}, {Fardal},
  {Mackey}, {Navarro}, {Pe{\~n}arrubia}, {Rich}, {Tanvir}, \&
  {Widrow}}]{ibata14}
{Ibata}, R.~A., {Lewis}, G.~F., {McConnachie}, A.~W., et al. 2014,
  \apj, 780, 128

\bibitem[{{Johnston} {et~al.}(2008){Johnston}, {Bullock}, {Sharma}, {Font},
  {Robertson}, \& {Leitner}}]{johnston08}
{Johnston}, K.~V., {Bullock}, J.~S., {Sharma}, S., et al. 2008, \apj, 689, 936

\bibitem[{{Kazantzidis} {et~al.}(2008){Kazantzidis}, {Bullock}, {Zentner},
  {Kravtsov}, \& {Moustakas}}]{kazantzidis08}
{Kazantzidis}, S., {Bullock}, J.~S., {Zentner}, A.~R., {Kravtsov}, A.~V., \&
  {Moustakas}, L.~A. 2008, \apj, 688, 254

\bibitem[{{Kordopatis} {et~al.}(2013){Kordopatis}, {Gilmore}, {Wyse},
  {Steinmetz}, {Siebert}, {Bienaym{\'e}}, {McMillan}, {Minchev}, {Zwitter},
  {Gibson}, {Seabroke}, {Grebel}, {Bland-Hawthorn}, {Boeche}, {Freeman},
  {Munari}, {Navarro}, {Parker}, {Reid}, \& {Siviero}}]{kordopatis13}
{Kordopatis}, G., {Gilmore}, G., {Wyse}, R.~F.~G., et al. 2013, \mnras, 436, 3231

\bibitem[{{Majewski} {et~al.}(2004){Majewski}, {Ostheimer}, {Rocha-Pinto},
  {Patterson}, {Guhathakurta}, \& {Reitzel}}]{majewski04b}
{Majewski}, S.~R., {Ostheimer}, J.~C., {Rocha-Pinto}, H.~J., et al. 2004, \apj, 615, 738

\bibitem[{{Martin} {et~al.}(2013{\natexlab{a}}){Martin}, {Carlin}, {Newberg},
  \& {Grillmair}}]{martinc13}
{Martin}, C., {Carlin}, J.~L., {Newberg}, H.~J., \& {Grillmair}, C.
  2013{\natexlab{a}}, \apjl, 765, L39

\bibitem[{{Martin} {et~al.}(2008){Martin}, {de Jong}, \& {Rix}}]{martin08b}
{Martin}, N.~F., {de Jong}, J.~T.~A., \& {Rix}, H.-W. 2008, \apj, 684, 1075

\bibitem[{{Martin} {et~al.}(2007{\natexlab{a}}){Martin}, {Ibata}, {Chapman},
  {Irwin}, \& {Lewis}}]{martin07a}
{Martin}, N.~F., {Ibata}, R.~A., {Chapman}, S.~C., {Irwin}, M., \& {Lewis},
  G.~F. 2007{\natexlab{a}}, \mnras, 380, 281

\bibitem[{{Martin} {et~al.}(2007{\natexlab{b}}){Martin}, {Ibata}, \&
  {Irwin}}]{martin07b}
{Martin}, N.~F., {Ibata}, R.~A., \& {Irwin}, M. 2007{\natexlab{b}}, \apjl, 668,
  L123

\bibitem[{{Martin} {et~al.}(2013{\natexlab{b}}){Martin}, {Ibata},
  {McConnachie}, {Mackey}, {Ferguson}, {Irwin}, {Lewis}, \&
  {Fardal}}]{martin13b}
{Martin}, N.~F., {Ibata}, R.~A., {McConnachie}, A.~W., et al.
  2013{\natexlab{b}}, \apj, 776, 80

\bibitem[{{Mart{\'{\i}}nez-Delgado} {et~al.}(2010){Mart{\'{\i}}nez-Delgado},
  {Gabany}, {Crawford}, {Zibetti}, {Majewski}, {Rix}, {Fliri},
  {Carballo-Bello}, {Bardalez-Gagliuffi}, {Pe{\~n}arrubia}, {Chonis}, {Madore},
  {Trujillo}, {Schirmer}, \& {McDavid}}]{martinez-delgado10}
{Mart{\'{\i}}nez-Delgado}, D., {Gabany}, R.~J., {Crawford}, K., et al. 2010, \aj, 140, 962

\bibitem[{{Mart{\'{\i}}nez-Delgado} {et~al.}(2008){Mart{\'{\i}}nez-Delgado},
  {Pe{\~n}arrubia}, {Gabany}, {Trujillo}, {Majewski}, \&
  {Pohlen}}]{martinez-delgado08}
{Mart{\'{\i}}nez-Delgado}, D., {Pe{\~n}arrubia}, J., {Gabany}, R.~J., et al. 2008, \apj, 689, 184

\bibitem[{{McConnachie} {et~al.}(2008){McConnachie}, {Huxor}, {Martin},
  {Irwin}, {Chapman}, {Fahlman}, {Ferguson}, {Ibata}, {Lewis}, {Richer}, \&
  {Tanvir}}]{mcconnachie08}
{McConnachie}, A.~W., {Huxor}, A., {Martin}, N.~F., et al. 2008, \apj, 688, 1009

\bibitem[{{McConnachie} {et~al.}(2009){McConnachie}, {Irwin}, {Ibata},
  {Dubinski}, {Widrow}, {Martin}, {C{\^o}t{\'e}}, {Dotter}, {Navarro},
  {Ferguson}, {Puzia}, {Lewis}, {Babul}, {Barmby}, {Bienaym{\'e}}, {Chapman},
  {Cockcroft}, {Collins}, {Fardal}, {Harris}, {Huxor}, {Mackey},
  {Pe{\~n}arrubia}, {Rich}, {Richer}, {Siebert}, {Tanvir}, {Valls-Gabaud}, \&
  {Venn}}]{mcconnachie09}
{McConnachie}, A.~W., {Irwin}, M.~J., {Ibata}, et al. 2009, \nat, 461, 66

\bibitem[{{Mouhcine} {et~al.}(2010){Mouhcine}, {Ibata}, \&
  {Rejkuba}}]{mouhcine10}
{Mouhcine}, M., {Ibata}, R., \& {Rejkuba}, M. 2010, \apjl, 714, L12

\bibitem[{{Pawlowski} {et~al.}(2013){Pawlowski}, {Kroupa}, \&
  {Jerjen}}]{pawlowski13a}
{Pawlowski}, M.~S., {Kroupa}, P., \& {Jerjen}, H. 2013, \mnras, 435, 1928

\bibitem[{{Pe{\~n}arrubia} {et~al.}(2002){Pe{\~n}arrubia}, {Kroupa}, \&
  {Boily}}]{penarrubia02}
{Pe{\~n}arrubia}, J., {Kroupa}, P., \& {Boily}, C.~M. 2002, \mnras, 333, 779

\bibitem[{{Pe{\~n}arrubia} {et~al.}(2005){Pe{\~n}arrubia},
  {Mart{\'{\i}}nez-Delgado}, {Rix}, {G{\'o}mez-Flechoso}, {Munn}, {Newberg},
  {Bell}, {Yanny}, {Zucker}, \& {Grebel}}]{penarrubia05}
{Pe{\~n}arrubia}, J., {Mart{\'{\i}}nez-Delgado}, D., {Rix}, H.~W.,
 et al. 2005, \apj, 626, 128

\bibitem[{{Pe{\~n}arrubia} {et~al.}(2006){Pe{\~n}arrubia}, {McConnachie}, \&
  {Babul}}]{penarrubia06}
{Pe{\~n}arrubia}, J., {McConnachie}, A., \& {Babul}, A. 2006, \apjl, 650, L33

\bibitem[{{Pila-D{\'{\i}}ez} {et~al.}(2013){Pila-D{\'{\i}}ez}, {Kuijken}, {de
  Jong}, {Hoekstra}, \& {van der Burg}}]{pila-diez14}
{Pila-D{\'{\i}}ez}, B., {Kuijken}, K., {de Jong}, J.~T.~A., {Hoekstra}, H., \&
  {van der Burg}, R.~F.~J. 2013, ArXiv:1311.7580

\bibitem[{{Purcell} {et~al.}(2011){Purcell}, {Bullock}, {Tollerud}, {Rocha}, \&
  {Chakrabarti}}]{purcell11}
{Purcell}, C.~W., {Bullock}, J.~S., {Tollerud}, E.~J., {Rocha}, M., \&
  {Chakrabarti}, S. 2011, \nat, 477, 301

\bibitem[{{Richardson} {et~al.}(2008){Richardson}, {Ferguson}, {Johnson},
  {Irwin}, {Tanvir}, {Faria}, {Ibata}, {Johnston}, {Lewis}, {McConnachie}, \&
  {Chapman}}]{richardson08}
{Richardson}, J.~C., {Ferguson}, A.~M.~N., {Johnson}, R.~A., et al. 2008, \aj, 135, 1998

\bibitem[{{Robin} {et~al.}(2007){Robin}, {Rich}, {Aussel}, {Capak}, {Tasca},
  {Jahnke}, {Kakazu}, {Kneib}, {Koekemoer}, {Leauthaud}, {Lilly}, {Mobasher},
  {Scoville}, {Taniguchi}, \& {Thompson}}]{robin07}
{Robin}, A.~C., {Rich}, R.~M., {Aussel}, H., et al. 2007, \apjs, 172, 545

\bibitem[{{Rocha-Pinto} {et~al.}(2004){Rocha-Pinto}, {Majewski}, {Skrutskie},
  {Crane}, \& {Patterson}}]{rocha-pinto04}
{Rocha-Pinto}, H.~J., {Majewski}, S.~R., {Skrutskie}, M.~F., {Crane}, J.~D., \&
  {Patterson}, R.~J. 2004, \apj, 615, 732

\bibitem[{{Siebert} {et~al.}(2012){Siebert}, {Famaey}, {Binney}, {Burnett},
  {Faure}, {Minchev}, {Williams}, {Bienaym{\'e}}, {Bland-Hawthorn}, {Boeche},
  {Gibson}, {Grebel}, {Helmi}, {Just}, {Munari}, {Navarro}, {Parker}, {Reid},
  {Seabroke}, {Siviero}, {Steinmetz}, \& {Zwitter}}]{siebert12}
{Siebert}, A., {Famaey}, B., {Binney}, et al. 2012, \mnras, 425, 2335

\bibitem[{{Sollima} {et~al.}(2011){Sollima}, {Valls-Gabaud},
  {Martinez-Delgado}, {Fliri}, {Pe{\~n}arrubia}, \& {Hoekstra}}]{sollima11}
{Sollima}, A., {Valls-Gabaud}, D., {Martinez-Delgado}, et al. 2011, \apjl, 730, L6

\bibitem[{{Taylor} \& {Babul}(2001)}]{taylor01}
{Taylor}, J.~E., \& {Babul}, A. 2001, \apj, 559, 716

\bibitem[{{Tollerud} {et~al.}(2012){Tollerud}, {Beaton}, {Geha}, {Bullock},
  {Guhathakurta}, {Kalirai}, {Majewski}, {Kirby}, {Gilbert}, {Yniguez},
  {Patterson}, {Ostheimer}, {Cooke}, {Dorman}, {Choudhury}, \&
  {Cooper}}]{tollerud12}
{Tollerud}, E.~J., {Beaton}, R.~L., {Geha}, M.~C., et al. 2012, \apj,
  752, 45

\bibitem[{{Widrow} {et~al.}(2012){Widrow}, {Gardner}, {Yanny}, {Dodelson}, \&
  {Chen}}]{widrow12}
{Widrow}, L.~M., {Gardner}, S., {Yanny}, B., {Dodelson}, S., \& {Chen}, H.-Y.
  2012, \apjl, 750, L41

\end{thebibliography}

\clearpage
\clearpage

\end{document}